\documentclass[twocolumn,preprintnumbers,aps,pra,superscriptaddress,amsmath,amssymb]{revtex4-1}

\usepackage{graphicx}
\usepackage{dcolumn}
\usepackage{bm}
\usepackage{bbm}
\usepackage{hyperref}
\usepackage{color}
\usepackage{multirow}

\begin{document}

\title{Microwave-free nuclear magnetic resonance at molecular scales}
	
\author{James D. A. Wood}\email{james.wood@unibas.ch}
\altaffiliation{Present address: Department of Physics, University of Basel, Switzerland}
\affiliation{Centre for Quantum Computation and Communication Technology, School of Physics, The University of Melbourne, VIC 3010, Australia}

\author{Jean-Philippe Tetienne}\email{jtetienne@unimelb.edu.au}
\affiliation{Centre for Quantum Computation and Communication Technology, School of Physics, The University of Melbourne, VIC 3010, Australia}

\author{David A. Broadway}
\affiliation{Centre for Quantum Computation and Communication Technology, School of Physics, The University of Melbourne, VIC 3010, Australia}

\author{Liam T. Hall}
\affiliation{School of Physics, The University of Melbourne, VIC 3010, Australia}

\author{David A. Simpson}
\affiliation{School of Physics, The University of Melbourne, VIC 3010, Australia}
	
\author{Alastair Stacey}
\affiliation{Centre for Quantum Computation and Communication Technology, School of Physics, The University of Melbourne, VIC 3010, Australia}
\affiliation{Melbourne Centre for Nanofabrication, 151 Wellington Road, Clayton, VIC 3168, Australia}

\author{Lloyd C. L. Hollenberg}
\affiliation{Centre for Quantum Computation and Communication Technology, School of Physics, The University of Melbourne, VIC 3010, Australia}
\affiliation{School of Physics, The University of Melbourne, VIC 3010, Australia}

\begin{abstract}
The implementation of nuclear magnetic resonance (NMR) at the nanoscale is a major challenge, as conventional systems require relatively large ensembles of spins and limit resolution to mesoscopic scales. New approaches based on quantum spin probes, such as the nitrogen-vacancy (NV) centre in diamond, have recently achieved nano-NMR under ambient conditions. However, the measurement protocols require application of complex microwave pulse sequences of high precision and relatively high power, placing limitations on the design and scalability of these techniques. Here we demonstrate a microwave-free method for nanoscale NMR using the NV centre, which is a far less invasive, and vastly simpler measurement protocol. By utilising a carefully tuned magnetic cross-relaxation interaction between a subsurface NV spin and an external, organic environment of proton spins, we demonstrate NMR spectroscopy of $^1$H within a $\approx(10~{\rm nm})^3$ sensing volume. We also theoretically and experimentally show that the sensitivity of our approach matches that of existing microwave control-based techniques using the NV centre. Removing the requirement for coherent manipulation of either the NV or the environmental spin quantum states represents a significant step towards the development of robust, non-invasive nanoscale NMR probes.
\end{abstract}	

\maketitle

\section{Introduction}

The discovery of nuclear magnetic resonance (NMR), and its related technologies, was one of the great scientific achievements of the 20th century, contributing to significant advances in areas ranging from materials science to healthcare. However, the limitation in sensitivity of traditional induction-based detection has required the development of new methods in order to extend NMR technology to the nanoscale, where the study of processes at the molecular scale is of intense interest. The nitrogen-vacancy (NV) centre in diamond \cite{Doherty2013} has seen remarkable developments as a high sensitivity nanoscale magnetometer \cite{Chernobrod2005,Degen2008,Taylor2008,Balasubramanian2008,Maze2008,Cole2009,Hall2009,Rondin2014,Schirhagl2014}. In recent years, the NV centre has been utilised to achieve nanoscale NMR at sensitivities close to that required for single proton detection \cite{Mamin2013,Staudacher2013,Loretz2014,Muller2014,DeVience2015,Lovchinsky2016}. Its room-temperature operation also makes it an ideal candidate for biological nano-NMR \cite{McGuinness2011,Shi2015}. However, current NMR protocols using the NV centre require the application of complex, high-power and high-precision microwave pulsing sequences in order to filter the environmental spectrum, placing significant constraints on nano-magnetic resonance imaging applications \cite{Loretz2015}. The application of strong microwave pulses is potentially invasive given the attendant electric fields as an inevitable by-product in the generation of the magnetic control fields. In addition, achieving such quantum control on a large ensemble of NV centres over a wide field of view \cite{Simpson2016}, or on a scanning probe microscopy tip \cite{Maletinsky2012,Tetienne2014,Tetienne2016}, remains a challenge due to the requirement for homogeneity in the driving field. Finally, the adaptability of such high-precision control to dynamic environments such as in vitro \cite{McGuinness2011} is unknown.

\begin{figure}[t]
\begin{center}
\includegraphics[width=0.48\textwidth]{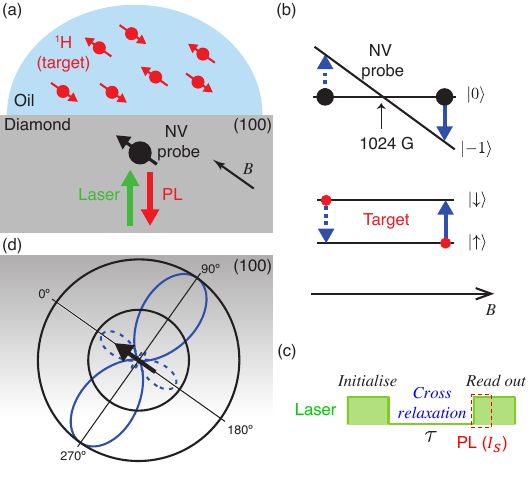}
\caption{Principle of microwave-free nano-NMR. (a) Schematic of the experimental setup. The protons within an organic sample external to the diamond are the targets probed by a shallow NV centre. A 532 nm green laser is used to initialise the NV spin while the red photoluminescence (PL) is measured via a single-photon detector. The background magnetic field is aligned with the NV quantisation axis with a variable strength, $B$. (b) Schematic of the energy levels of the NV (spin states $|0\rangle$ and $|-1\rangle$, neglecting the hyperfine structure for clarity) and of a target nuclear spin such as $^{1}$H (states $|\uparrow\rangle$ and $|\downarrow\rangle$). As $B$ is swept across the GSLAC, two resonances can in principle occur where the NV and target spins can exchange energy via their mutual magnetic dipole-dipole interaction, causing an increase in their respective longitudinal relaxation rate. (c) The measurement sequence consists of laser pulses to initialise and subsequently read out the NV spin state, separated by a wait time $\tau$. (d) Polar plot of the relaxation rate induced by a single spin, $\Gamma_{1,{\rm ext}}$, as a function of the angle $\theta$ between the quantisation axis and the NV-target separation. The solid (dashed) line corresponds to the after-GSLAC (before-GSLAC) resonance. The values are normalised by the global maximum, so that the outer circle corresponds to maximum strength.} 
\label{Fig1}
\end{center}
\end{figure}

In this work, we demonstrate microwave-free nano-NMR on a nanoscopic sample of proton nuclear spins within an organic sample external to the diamond (Fig. \ref{Fig1}a).  We achieve this via static field tuning to the natural spin interactions between sensor and target, and measure a change in the longitudinal relaxation time, $T_1$ \cite{Hall2016}. The resulting all-optical microwave-free protocol can be applied in a non-invasive manner (e.g., through optical excitation constrained within the diamond using total internal reflection \cite{Clevenson2015}), significantly widening the range of applications where NV-based nano-NMR can be used. We demonstrate the microwave-free NMR technique using both isotopic forms $^{14}$NV and $^{15}$NV, which exhibit different fundamental characteristics in this context, and use the data to estimate the NV-sample distance ($10-12$~nm for the studied NVs) as well as the number of protons detected ($<10^5$).  In addition, we directly compare our microwave-free nano-NMR to the prevailing microwave control based nano-NMR technique, and find that both approaches offer comparable sensitivity.

\section{Principle of $T_1$-NMR}

The microwave-free technique for magnetic resonance spectroscopy is based on $T_1$ relaxometry detection \cite{Cole2009,Kaufmann2013,Steinert2013,Tetienne2013,Sushkov2014} and precise magnetic field tuning to bring the NV into resonance with the Zeeman split target spin transitions \cite{Hall2016}. This technique was proposed and demonstrated for electron spin resonance spectroscopy using an ensemble of NV centres \cite{Hall2016} and extended to hyperfine-coupled nuclear spin spectroscopy with single NV centres \cite{Wood2016}. The ultimate goal of nano-NMR is the spectroscopic detection of bare nuclear spins in molecular systems. Without the assistance of strong hyperfine coupling this is a significant challenge due to the low nuclear magnetic moment \cite{Wood2016,Broadway2016}.  Applying the $T_1$-NMR technique to this problem requires the precise tuning of the NV probe to near its ground state level anti-crossing (GSLAC), where the system's dynamics are dominated by a relatively complex landscape of electron-nuclear spin mixing \cite{He1993,Epstein2005,Broadway2016}, in order to bring the electronic NV transitions into resonance with the nuclear spin  transition. At a resonance point (which in principle could occur both before and after the GSLAC as depicted in Fig. \ref{Fig1}b), the longitudinal relaxation time $T_1$ of both the NV centre and the environmental spin(s) are significantly reduced, due to their mutual dipole-dipole interaction \cite{Hall2016,Wood2016,Jarmola2012}. In order to probe these resonances, the strength, $B$, of a static background magnetic field aligned with the NV quantisation axis, is swept near the GSLAC ($B\approx 1024$~G) and the $T_1$ decay is optically measured at each corresponding transition frequency. 

The transitions associated with two simple resonances before and after the GSLAC are depicted in Fig.~\ref{Fig1}b. The relaxation rate of the initialised NV spin state $|0\rangle$ to $|-1\rangle$, induced at one of these resonances (over and above the intrinsic relaxation rate $\Gamma_{\rm 1,int}$, assumed constant in the field range considered), is set by the sum of the contributions of all resonant target spins, $\Gamma_{1,{\rm ext}}^\pm=\sum_i\Gamma_{1,{\rm ext}}^{i\pm}$. The contribution of each $i$th spin can be expressed as \cite{Wood2016}
\begin{equation} \label{eq:GammaInt+-}
\Gamma_{1,{\rm ext}}^{i\pm} = \frac{1}{2\Gamma_{2}^*}\left(\frac{\mu_0\gamma_{\rm NV}\gamma_{\rm t} \hbar}{4\pi} \right)^2 \left(\frac{3\sin^2\theta_i-1\pm1}{r_i^3} \right)^2
\end{equation}
where $\gamma_{\rm NV}$ and $\gamma_{\rm t}$ are the gyromagnetic ratios of the NV and target spins, respectively; $\theta_i$ is the polar angle of separation between the two spins relative to the quantisation axis; $r_i$ is the separation distance; $\Gamma_{2}^*$ is the total dephasing rate of the spin system; $\mu_0$ is the vacuum permeability; $\hbar$ is Planck's constant; and the $\pm$ sign refers to the resonance after ($+$) and before ($-$) the GSLAC. Near resonance, the photoluminescence (PL) signal after an evolution time $\tau$ is given by \cite{Steinert2013,Kaufmann2013,SI}  
\begin{equation} \label{eq:PLfunction}
I_s(\tau,B)=I_0\left[1+{\cal C}e^{-\Gamma_{1,\rm tot}(B)\tau}\right],
\end{equation}
where $I_0$ is a normalisation constant, ${\cal C}\sim0.2$ is the PL contrast of the $T_1$ decay, and the total relaxation rate is \cite{Hall2016}
\begin{equation} \label{eq:GammaTot}
\Gamma_{1,\rm tot}(B)=\Gamma_{1,\rm int}+\frac{\Gamma_{1,\rm ext}^\pm}{1+\left(\frac{\omega_{\rm NV}(B)-\omega_{\rm t}(B)}{\Gamma_2^*}\right)^2},
\end{equation}
with $\omega_{\rm NV}$ the NV transition frequency and $\omega_{\rm t}=\gamma_{\rm t}B$ the Larmor frequency of the target spin, which vary with $B$ as depicted in Fig. \ref{Fig1}b. Eq. (\ref{eq:PLfunction}) is valid when the intrinsic relaxation ($\Gamma_{1,\rm int}$) is dominated by magnetic noise acting on the $|0\rangle\leftrightarrow|-1\rangle$ transition, which is generally the case near the GSLAC \cite{SI}. Experimentally, rather than recording a full curve $I_s(\tau)$, it is generally sufficient to probe a single, well chosen value of $\tau$, from which one can retrieve $\Gamma_{1,\rm tot}(B)$ using Eq. (\ref{eq:PLfunction}), provided $I_0$ and ${\cal C}$ are initially calibrated via measurement of a full decay trace $I_s(\tau,B)$ at a given $B$. Such a single-point $T_1$ measurement is done via an initialise, wait, and read-out scheme (Fig. \ref{Fig1}c), without any microwave pulsing of either the NV spin or target spin. We note that the relaxation rate induced by each target spin, expressed by Eq. (\ref{eq:GammaInt+-}), has a different overall angular dependence for the two resonances before and after the GSLAC, as illustrated in Fig. \ref{Fig1}d. This difference allows $\theta$ to be inferred, in the case of a single target spin, by directly comparing the transition strengths on either side of the GSLAC.

The simplified picture of resonances near the GSLAC shown in Fig. \ref{Fig1}b is in reality modified by hyperfine interactions within the NV centre, associated with the nitrogen nuclear spin which is either $^{14}$N (spin-1) or $^{15}$N (spin-1/2) \cite{He1993,Epstein2005,Broadway2016}. While sensing experiments generally use implanted $^{15}$NV centres, as they can be distinguished from native centres of the most abundant isotope, $^{14}$N \cite{Rabeau2006}, it has become apparent from recent experiments \cite{Broadway2016} that $^{14}$NV and $^{15}$NV centres present distinct differences when considering their application for $T_1$-NMR spectroscopy due to their different GSLAC electron-nuclear state mixing properties \cite{Broadway2016}. For this reason, both isotopic forms of the NV probe were investigated in this work.

\section{Nano-NMR detection of $^1$H spins}

To demonstrate $T_1$-based NMR, a custom-built confocal microscope was used, incorporating a permanent magnet mounted on a 3-axis scanning stage to precisely control the applied background field \cite{Wood2016}. The diamond sample comprises a high purity CVD homoepitaxial layer, grown in a Seki AX6500 diamond reactor. The sample was implanted firstly with 3.5 keV $^{15}N^+$ ions and then with 3.5 keV  $^{14}N^+$ ions (expected implantation depth in the range $5-15$~nm \cite{Lehtinen2016}). Both implants were done to a dose of $0.5 \times 10^9$ cm$^{-2}$ and this was followed by annealing in vacuum at $800^\circ$C and acid cleaning in a boiling mixture of sulphuric acid and sodium nitrate. The diamond has a (100) surface, so that the NV centres have their symmetry axis at $54.7^\circ$ to the surface normal (see Figs. \ref{Fig1}a and \ref{Fig1}d). The NV centres were identified as either $^{15}$NV or $^{14}$NV via their characteristic hyperfine splittings under low-power optically detected magnetic resonance (ODMR) \cite{Rabeau2006,Dreau2011}. As a standard proton sample, we used either the oil employed with the oil-immersion objective lens, or a layer of Poly(methyl methacrylate) (PMMA) deposited on the diamond surface \cite{Staudacher2013,Mamin2013}.

\begin{figure}[t]
\begin{center}
\includegraphics[width=0.48\textwidth]{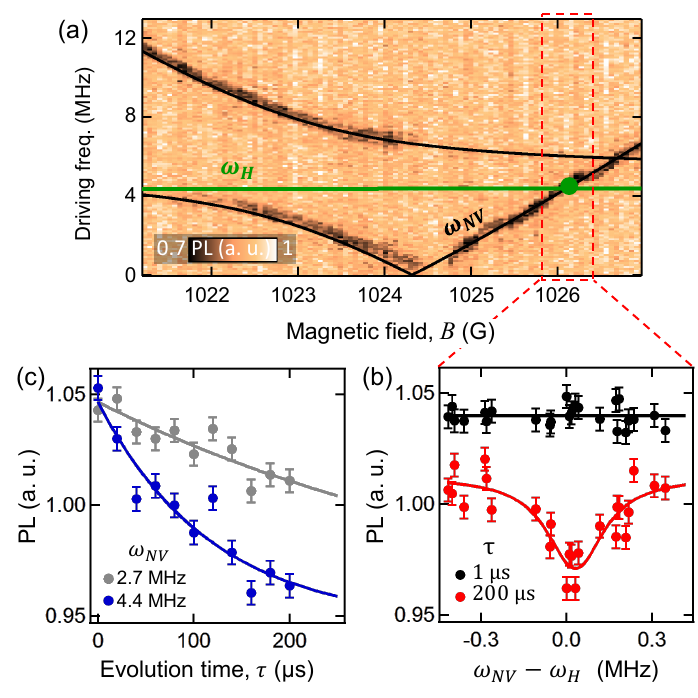}
\caption{Nano-NMR detection of $^1$H spins using $^{14}$NV. (a)~ODMR spectra of a $^{14}$NV centre measured as a function of the axial magnetic field strength, $B$, near the GSLAC. Solid lines overlaid on the graph are the calculated NV (black lines) and $^1$H (green line) spin transitions. (b) $T_1$-NMR spectrum around the $^1$H resonance, obtained with evolution times $\tau=1~\mu$s (black data) and $\tau=200~\mu$s (red data), plotted as a function of the detuning $\omega_{\rm NV}-\omega_{\rm H}$, where $\omega_{\rm NV}$ is obtained from fitting the ODMR spectrum and $\omega_{\rm H}$ is the $^1$H Larmor frequency. The PL is measured at the start of the readout pulse and normalised by the back of the same pulse. Solid lines are fits to Eqs. (\ref{eq:PLfunction}) and (\ref{eq:GammaTot}). (c) Full $T_1$ curves measured on resonance ($\omega_{\rm NV}=4.4$~MHz, blue data) and off resonance ($\omega_{\rm NV}=2.7$~MHz, grey data). Solid lines are fits to Eq. (\ref{eq:PLfunction}). Error bars represent the photon shot noise (one standard deviation).} 
\label{Fig2}
\end{center}
\end{figure}

We first report the results of $T_1$-NMR using a $^{14}$NV centre. The ODMR spectrum as a function of the background field strength ($B$) around 1024 G is shown in Fig. \ref{Fig2}a, revealing a characteristic GSLAC structure resulting from hyperfine couplings and dynamic polarisation of the NV nuclear spin \cite{Broadway2016,Jacques2009}. The NV transitions are shown as black lines, while the $^1$H spin transition is shown in green and corresponds to a Larmor frequency $\omega_{\rm H}\approx4.4$ MHz in this range of fields. Due to the NV avoided crossing before the GSLAC, there is only one measurable resonance point with $^1$H (green dot in Fig. \ref{Fig2}a). To probe this resonance, the PL intensity is measured as a function of $B$, using evolution times $\tau=1~\mu$s, which serves as reference for normalisation (as no $T_1$ decay is expected at such time scale), and $\tau=200~\mu$s, optimised to probe the $^1$H resonance. The resulting $T_1$-NMR spectrum is shown in Fig. \ref{Fig2}b, plotted against the detuning $\omega_{\rm NV}-\omega_{\rm H}$, where $\omega_{\rm NV}$ is the NV transition frequency obtained from fitting the ODMR spectrum. A dip in the $\tau=200~\mu$s data is clearly observed at the expected $^1$H frequency. Full $T_1$ curves measured on resonance ($\omega_{\rm NV}=4.4$~MHz) and off resonance ($\omega_{\rm NV}=2.7$~MHz) are shown in Fig. \ref{Fig2}c, confirming a change in relaxation time. Fitting the two curves to Eq. (\ref{eq:PLfunction}) yields the values $\Gamma_{1,\rm int}=2.1\pm0.2\times10^3$~s$^{-1}$ and $\Gamma_{1,\rm ext}^+=5.9\pm0.6\times10^3$~s$^{-1}$. The proton signal was confirmed to be associated primarily with the immersion oil, as the signal significantly reduced upon removal of the oil \cite{SI}.

\begin{figure}[t]
\begin{center}
\includegraphics[width=0.48\textwidth]{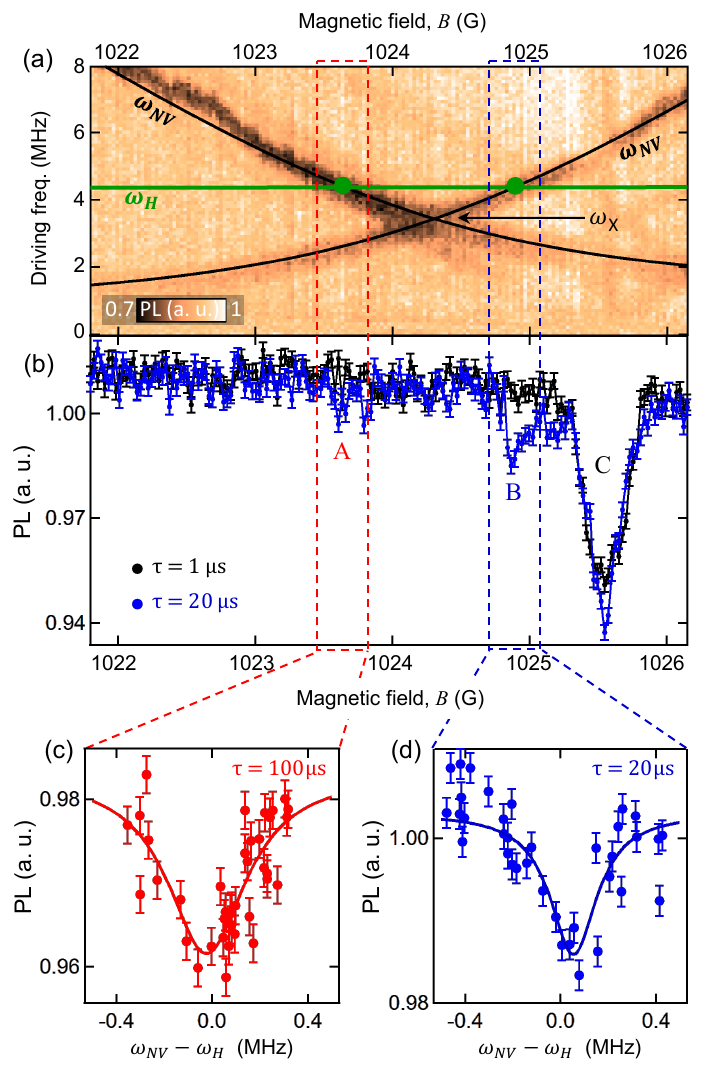}
\caption{Nano-NMR detection of $^1$H spins using $^{15}$NV. (a)~ODMR spectra of a $^{15}$NV centre measured as a function of $B$ near the GSLAC. The black lines are the calculated frequencies of the two dominant NV spin transitions, which cross at a frequency $\omega_\times$ (indicated by the black arrow). The frequency of the upper NV transition at each $B$ is denoted as $\omega_{\rm NV}$. The green line indicates the $^1$H transition frequency, $\omega_{\rm H}$. (b) $T_1$-NMR spectrum recorded with evolution times $\tau=1~\mu$s and $\tau=20~\mu$s. The three features observed are labelled A, B and C. (c,d) $T_1$-NMR spectra measured around feature A with $\tau=100~\mu$s (c) and around feature B with $\tau=20~\mu$s (d), plotted against $\omega_{\rm NV}-\omega_{\rm H}$, where $\omega_{\rm NV}$ is obtained from fitting the ODMR spectrum. Solid lines are fits to Eqs. (\ref{eq:PLfunction}) and (\ref{eq:GammaTot}). Error bars represent the photon shot noise (one standard deviation).}
\label{Fig3}
\end{center}
\end{figure}

We now demonstrate $T_1$-NMR using a $^{15}$NV centre from the same diamond and proton sample (immersion oil) as was used in the $^{14}$NV case. In Ref. \cite{Broadway2016}, it has been shown that a typical $^{15}$NV exhibits two potential resonances with $^1$H, before and after the GSLAC, but that the stronger after-GSLAC resonance overlaps with a feature intrinsic to the GSLAC structure. To circumvent this problem and allow the two resonances to be observed, we selected a $^{15}$NV centre exhibiting a hyperfine coupling (here of $\approx4$~MHz) to a nearby $^{13}$C, thereby shifting the intrinsic feature away from the NV-proton resonance \cite{Dreau2012,SI}. The results for this particular $^{15}$NV centre are shown in Fig. \ref{Fig3}. The ODMR spectrum as a function of $B$ around 1024 G is shown in Fig. \ref{Fig3}a, revealing the characteristic $^{15}$NV GSLAC structure \cite{Broadway2016}. The two dominant transitions (black lines) cross at a frequency $\omega_\times\approx3.5$~MHz, which is lower than the Larmor frequency of $^1$H in this range of fields ($\omega_{\rm H}$, green line). For each value of $B$, we define $\omega_{\rm NV}$ as the NV transition frequency of the upper branch, i.e. $\omega_{\rm NV}>\omega_\times$. The two expected $^1$H resonance points, $\omega_{\rm NV}=\omega_{\rm H}$, are shown as green dots in Fig. \ref{Fig3}a. A $T_1$-NMR spectrum recorded using evolution times $\tau=1~\mu$s (reference) and $\tau=20~\mu$s is shown in Fig. \ref{Fig3}b and reveals three main features, labelled A, B and C. The first two dips (A and B) are seen only for $\tau=20~\mu$s and correspond to the cross-relaxation resonances with $^1$H. The third dip, (C), exhibiting a far stronger decay, which is also visible at short time $\tau=1~\mu$s, corresponds to the intrinsic feature discussed above \cite{Broadway2016}. 

From Fig. \ref{Fig3}b, it can be seen that the lower-field $^1$H resonance (feature A) is significantly weaker than the higher-field resonance (feature B). To resolve this resonance more clearly, a longer evolution time, $\tau=100~\mu$s, was employed resulting in the spectrum shown in Fig. \ref{Fig3}c plotted against the detuning $\omega_{\rm NV}-\omega_{\rm H}$. Fitting the data to Eq. (\ref{eq:PLfunction}) along with Eq. (\ref{eq:GammaTot}) gives $\Gamma_{1,\rm int}=1.5\pm0.4\times10^3$~s$^{-1}$ and $\Gamma_{1,\rm ext}^-=1.5\pm0.3\times10^3$~s$^{-1}$ for this resonance. Similarly, the spectrum around the $^1$H resonance past the GSLAC (Fig. \ref{Fig3}d) is fitted to give $\Gamma_{1,\rm int}=1.3\pm0.2\times10^3$~s$^{-1}$ and $\Gamma_{1,\rm ext}^+=5.2\pm0.6\times10^3$~s$^{-1}$. The measured ratio between the extrinsic relaxation rates of the two resonances is $\frac{\Gamma_{1,{\rm ext}}^{+}}{\Gamma_{1,{\rm ext}}^{-}}\approx3.5$. This is in qualitative agreement with the ratio of 6.3 predicted by integrating Eq. (\ref{eq:GammaInt+-}) over a semi-infinite bath of protons on a (100) surface. This calculation neglects the effects of state mixing due to the GSLAC structure. 

\section{Comparison to $T_2$-based NMR}

We now report a theoretical and experimental comparison of $T_1$-NMR spectroscopy to the prevailing existing technique of $T_2$-based spectroscopy, which relies on locking a dynamical decoupling pulse sequence on the NV electronic spin state (usually XY8-$N$ where $N$ is the number of microwave $\pi$ pulses) to the nuclear spins' Larmor frequency \cite{Staudacher2013,Loretz2014,Muller2014,DeVience2015,Pham2015}. Theoretically, the sensitivity can be compared by examining the signal-to-noise ratio (SNR) derived under the identical conditions of a shallow NV centre detecting an ensemble of nuclear spins as in the geometry of Fig. \ref{Fig1}a. In the small signal regime (i.e., $\Gamma_{\alpha,\rm ext}\ll\Gamma_{\alpha,\rm int}$ with $\alpha=1$ for $T_1$ and $\alpha=2$ for $T_2$), we find that the maximum SNR in $T_1$ and $T_2$ sensing schemes can be expressed as \cite{SI}
\begin{eqnarray}
{\rm SNR}_{T_1,\rm max} & \approx & {\cal A}\times\frac{19\pi e^{-1/2}}{96\sqrt{2}}\times T_2^*\sqrt{T_{1}} \\
{\rm SNR}_{T_2,\rm max} & \approx & {\cal A}\times\frac{5(3/e)^{3/4}}{48\pi2^{3/2}}\times\left(T_{2}\right)^{3/2},
\end{eqnarray} 
respectively, where $T_2^*=1/\Gamma_2^*$, $T_1=1/\Gamma_{1,\rm int}$ and $T_2=1/\Gamma_{2,\rm int}$ are the intrinsic characteristic times of the NV centre. Note that here $T_2$ is the extended decoherence time under the considered dynamical decoupling sequence \cite{Pham2012}. The constant ${\cal A}$ is given by
\begin{eqnarray}
{\cal A}=\sqrt{{\cal R}T_{\rm tot}t_{\rm ro}}{\cal C}\frac{\rho}{d^3}\left(\frac{\mu_0\gamma_{\rm NV}\gamma_t\hbar}{4\pi}\right)^2
\end{eqnarray} 
where ${\cal R}$ is the photon count rate under continuous laser excitation, $t_{\rm ro}=300$ ns is the read-out time per pulse, $T_{\rm tot}$ is the total experiment time, $\rho$ is the proton density in the semi-infinite sample, and $d$ is the NV depth. Evaluating the numeric factors, we obtain the ratio between the SNRs,
\begin{eqnarray}
\frac{{\rm SNR}_{T_1,\rm max}}{{\rm SNR}_{T_2,\rm max}} & \approx & 21.1\times\frac{T_2^*\sqrt{T_1}}{(T_2)^{3/2}}.
\end{eqnarray} 
For a near-surface NV centre in a bulk diamond ($d\approx10$ nm), typical approximate values are $T_2^*=2~\mu$s, $T_1=2$~ms and $T_2=200~\mu$s, which yields a ratio $\frac{{\rm SNR}_{T_1,\rm max}}{{\rm SNR}_{T_2,\rm max}}\sim1$. In other words, our microwave-free $T_1$-based NMR spectroscopy technique is similarly sensitive to the existing $T_2$-based approach without requiring complex microwave pulse sequences.  

\begin{figure}[t]
\begin{center}
\includegraphics[width=0.48\textwidth]{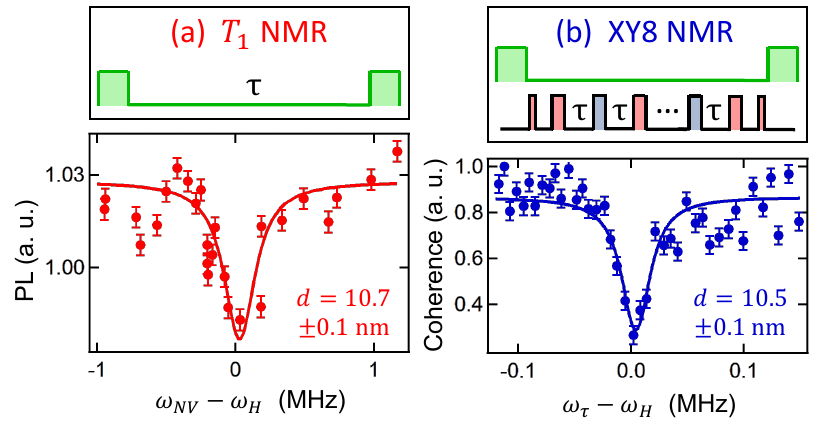}
\caption{Comparison of microwave-free $T_1$-NMR to the XY8 protocol. (a,b) Spectra from PMMA obtained with a single $^{14}$NV centre by using: (a) microwave-free $T_1$ relaxometry after the GSLAC, with a probe time $\tau=100~\mu$s; (b) an XY8-$N$ dynamical decoupling sequence with $N=256$ microwave pulses at a field $B=300$~G. The corresponding pulse sequence is depicted above each graph, with laser pulses in green and microwave pulses in red or blue corresponding to $0^\circ$ or $90^\circ$ relative phase. In (a), the spectrum is constructed by varying the NV frequency, $\omega_{\rm NV}$, via varying the magnetic field strength. In (b), it is constructed by scanning the probe frequency $\omega_\tau=\pi/\tau$, where $\tau$ is the inter-pulse delay, inclusive of the finite $\pi$ pulse duration. Error bars represent the photon shot noise (one standard deviation).}
\label{Fig4}
\end{center}
\end{figure}

In order to verify this and further compare the two approaches, we conducted a comparative measurement using a single shallow $^{14}$NV centre, where the diamond was coated with a layer of PMMA. Fig. \ref{Fig4} shows the hydrogen spectrum measured via $T_1$ relaxometry (a) and via an XY8-256 sequence (b). Both spectra were acquired in a total time of about 2 hours, and show a clear feature at the $^1$H frequency with a similar signal-to-noise ratio. In addition, one can compare the NV depth, $d$, inferred through fitting the appropriately normalised data from each method, assuming the signal comes from a semi-infinite bath of protons on a (100) surface \cite{SI}. Using a proton density of $\rho=56~{\rm nm}^{-3}$, we find $d=10.7\pm0.1$~nm from the $T_1$ data, against $d=10.5\pm0.1$~nm from the XY8 data, indicating a high level of consistency between the two approaches. From the inferred depth, we deduce that 50\% of the signal is generated by the $\approx6\times10^4$ closest protons, corresponding to a detection volume of about $(10~{\rm nm})^3$ \cite{Staudacher2013,Loretz2014}.

The spectral resolution of $T_1$ spectroscopy, however, is currently limited by the dephasing rate $\Gamma_2^*=1/T_2^*$ (see Eq. (\ref{eq:GammaTot})), while it is limited by $T_2$ with the XY8 method, and can be improved further using correlation spectroscopy \cite{Staudacher2015,Kong2015,Boss2016}. We note that both the sensitivity and spectral resolution of the $T_1$ approach could be dramatically improved by optimising $T_2^*$, which motivates further work towards understanding and mitigating the decoherence of near-surface NV centres \cite{Myers2014,Rosskopf2014,Romach2015}.

\section{Conclusions}

In summary, we have demonstrated the detection of proton spins external to a diamond using a microwave-free nano-NMR technique based on $T_1$ relaxometry of a single NV centre. While the electron-nuclear spin physics near the NV GSLAC are relatively complex, we have shown that the microwave-free $T_1$ protocol can nevertheless be implemented and the $^1$H resonances observed, which we have demonstrated using both isotopic forms of the NV centre. In addition, we have shown a sensitivity comparable to an existing nano-NMR protocol which requires quantum state manipulation via microwave excitation. The sensitivity as well as the spectral resolution of our approach could be improved via engineering of the diamond surface. We note that the resulting all-optical technique can be implemented using non-invasive excitation constrained within the diamond, using total internal reflection. The removal of microwave quantum control eliminates the possibility of spurious harmonics within the measurement and opens up applications in areas where such control is inherently difficult to achieve and/or invasive, such as spectroscopic imaging over wide fields of view or in scanning probe microscopy experiments.
\\

\section*{Acknowledgements}

This work was supported in part by the Australian Research Council (ARC) under the Centre of Excellence scheme (project No. CE110001027). L.C.L.H. acknowledges the support of an ARC Laureate Fellowship (project No. FL130100119). This work was supported in part by the Melbourne Centre for Nanofabrication (MCN) in the Victorian Node of the Australian National Fabrication Facility (ANFF).

\end{document}